\documentclass[iop]{emulateapj}
\usepackage{apjfonts}
\submitted{Received 2011 February 3; accepted 2011 May 16; published --}
\journalinfo{Accepted to ApJL}
\shorttitle{A STANDARD-TO-BLOWOUT JET}
\shortauthors{LIU ET AL.}
\newcommand{\ha}{H$\alpha$}

\newcommand{\Hsi}{\textit{Reuven Ramaty High Energy Solar Spectroscopic Imager}}
\newcommand{\hsi}{\textit{RHESSI}}
\newcommand{\goes}{\textit{GOES}}
\newcommand{\sm}{$\sim$}
\newcommand{\Sdo}{\textit{Solar Dynamics Observatory}}
\newcommand{\sdo}{\textit{SDO}}
\newcommand{\hmi}{Helioseismic and Magnetic Imager}
\newcommand{\aia}{Atmospheric Imaging Assembly}
\newcommand{\stereo}{\textit{STEREO}}

\begin{document}

\title{A STANDARD-TO-BLOWOUT JET}
\author{Chang Liu\altaffilmark{1}, Na Deng\altaffilmark{2,1}, Rui Liu\altaffilmark{1}, Ignacio Ugarte-Urra\altaffilmark{3}, Shuo Wang\altaffilmark{1}, and Haimin Wang\altaffilmark{1}}
\affil{1. Space Weather Research Laboratory, New Jersey Institute of Technology, University Heights, Newark, NJ 07102-1982, USA; chang.liu@njit.edu}
\affil{2. Department of Physics and Astronomy, California State University, Northridge, CA 91330-8268, USA}
\affil{3. College of Science, George Mason University, 4400 University Drive, Fairfax, VA 22030-4422, USA}

\begin{abstract}
The commonly observed jets provide critical information on the small-scale energy release in the solar atmosphere. We report a near disk-center jet on 2010 July 20, observed by the \Sdo. In this event, the standard interchange magnetic reconnection between an emerging flux spanning 9$\times10^3$~km and ambient open fields is followed by a blowout-like eruption. In the ``standard'' stage, as the emerging negative element approached the nearby positive network fields, a jet with a dome-like base in EUV grew for 30~minutes before the jet spire began to migrate laterally with enhanced flux emergence. In the ``blowout'' stage, the above converging fields collided and the subsequent cancellation produced a UV microflare lasting seven minutes, in which the dome of the jet seemed to be blown out as (1) the spire swung faster and exhibited an unwinding motion before vanishing, (2) a rising loop and a blob erupted leaving behind cusped structures, with the blob spiraling outward in acceleration after the flare maximum, and (3) ejecting material with a curtain-like structure at chromospheric to transition-region temperatures also underwent a transverse motion. It is thus suggested that the flare reconnection rapidly removes the outer fields of the emerging flux to allow its twisted core field to erupt, a scenario favoring the jet-scale magnetic breakout model as recently advocated by Moore et al. in 2010.
\end{abstract}


\section{INTRODUCTION} \label{introduction}
It is believed that jet activities in the solar atmosphere is commonly associated with flux emergence and cancellation \citep[e.g.,][]{chae99}. In the standard jet model \citep{shibata92}, a bipolar emerging flux intrudes into and continuously presses against the ambient high-reaching, opposite-polarity fields such as coronal hole regions with open field lines, which builds up a current sheet above the polarity inversion line (PIL). The subsequent interchange magnetic reconnection results in an inverted-Y shape of jets comprising (1) new open fields rooted in another leg of the emerging arch (called jet spire) along which heated plasmas is ejected outward, and (2) new closed fields across the PIL usually seen as bright points. This scenario has been demonstrated in two-dimensional and three-dimensional simulations \citep{yokoyama96,miyagoshi03,miyagoshi04,moreno08,archontis10,pariat10}, which also reproduce the dome-like magnetic topology of the jet base evidenced in observations.

By examining X-ray jets in polar coronal holes, a dichotomy of jets was recently put forward by \citet{moore10}. In this classification, about one third of jets are suggested to belong to a non-standard type, where the emerging magnetic arch that serves as the jet base is assumed to be sheared and twisted so that it can undergo a blowout eruption beneath the standard interchange reconnection, a process mimicking the breakout scenario for major coronal mass ejections (CMEs). Not only the signatures of blowout-like coronal jets \citep[][and references therein]{moore10} but also the counterpart in chromospheric spicules \citep{sterling10} have recently been observed. Simulations of blowout of the twisted jet core field by the breakout reconnection with ambient fields were also conducted \citep{pariat09,rachmeler10}.

An important aspect is that the twist possessed by the emerging flux can be transferred to the reconnected fields via the interchange reconnection. The escaping and relaxation of magnetic twist was proposed as a mechanism for the formation and acceleration of jets \citep{shibata85,shibata86} and was also demonstrated in recent jet simulations \citep{schmieder08,pariat09}. Signatures of upward twist propagation, such as unwinding motion and helical structure component, were reported for chromospheric spicules \citep{sterling10}, \ha\ surges \citep{canfield96}, polar coronal jets \citep{patsourakos08,liu+wei+jet09}, and coronal streamers \citep{rust09}.

Observational study of the jet dynamics has thus far been concentrated on coronal hole regions close to the limb. To better understand the physical process of jets, however, coronal observations combined with photospheric magnetic measurements are desired, especially in high-spatial and high-temporal resolution and with a wide temperature coverage. This has been made possible with the newly launched \Sdo\ (\sdo), the unprecedented observing capabilities of which gives a favorable opportunity to analyze the small-scale jets in detail.

In this Letter, we investigate a near disk-center jet on 2010 July 20 using \sdo\ data, in which flux emergence and cancellation are clearly seen as a dynamically related coronal jet experiences a standard-to-blowout transition. We also found unwinding and spinning jet features signifying the emergence and eruption of twisted magnetic fluxes.

\begin{figure*}
\epsscale{1.17}
\plotone{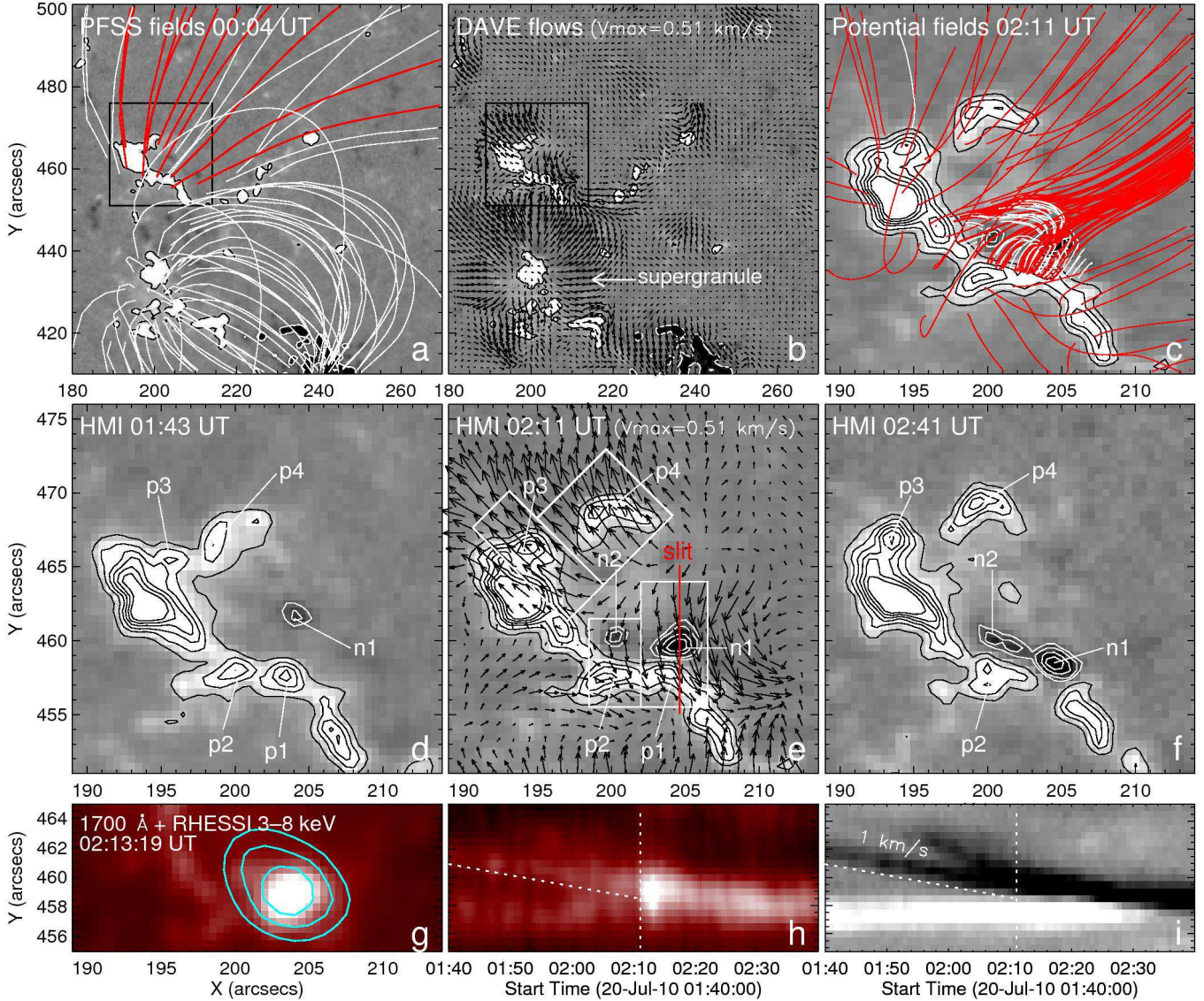}
\caption{Magnetic field topology and evolution ($a$--$f$) and a UV image at the flare peak ($g$). In $a$ and $c$, red and white lines represent open and closed fields, respectively. DAVE flows in $b$ and $e$ are averaged for 01:30--02:54~UT. The black box in $a$ and $b$ is the field-of-view (FOV) of $c$--$f$, where the levels of magnetic field contours are $-$300--$-$50 and 100--600~G. Contours (30\%, 50\%, 70\%, and 90\% of the maximum flux) in $g$ represent \hsi\ PIXON image integrated for 60~s centered at 02:13:19~UT. Time slices for the slit in $e$ using 1700~\AA\ and magnetogram images are shown in $h$ and $i$. \label{f1}}
\end{figure*}

\section{OBSERVATIONS AND DATA REDUCTION}
Line-of-sight magnetograms with a 45~s cadence and 1\arcsec\ resolution obtained with the \hmi\ \citep[HMI;][]{scherrer10} on board \sdo\ were used to trace the photospheric magnetic field evolution. The averaged flow fields was calculated using the differential affine velocity estimator \citep[DAVE;][]{schuck06}, the result of which was checked and confirmed by the local correlation tracking \citep{november88}. To reveal the magnetic topology of the source region, we resorted to both the full-Sun extrapolation using the potential field source surface \citep[PFSS;][]{schrijver03b} model based on MDI magnetograms and the potential field extrapolation in Cartesian coordinates using HMI magnetograms. The former provides a better context of the flux system of interest while lacking resolution to deal with small-scale structures \citep[e.g.,][]{ugarte07}.

The \aia\ \citep[AIA;][]{title10} on board \sdo\ takes images in seven EUV and three UV-visible wavelength bands. We take advantage of its observation in fine detail (0.6\arcsec~pixel$^{-1}$), high cadence (\sm12~s), and multiple wavelengths to examine the jet dynamics. The images were calibrated using the standard AIA procedure (\verb+aia_prep+) in the Solar SoftWare (SSW) and were co-registered, being differentially rotated to a reference time (2010 July 20 02:10~UT).

The jet-associated microflare, peaking at 02:13:30~UT, was registered by \goes\ 14 at A6 class of the incremental flux in 1--8~\AA\ and by the \Hsi\ \citep[\hsi;][]{lin02} up to \sm8~keV above the background. PIXON images \citep{hurford02} in the 3--8~keV energy range were reconstructed using the front segment of detectors 3--8. X-ray spectra in 20~s interval were analyzed using the Object Spectral Executive (OSPEX) package in SSW with $1/3$~keV energy bin width, the standard finest binning. In addition, radio dynamic spectra of this event were obtained by the Learmonth Solar Observatory, and SWAVES radio experiment \citep{bougeret08} on board \stereo\ spacecraft \citep{kaiser08}.

\section{RESULTS AND ANALYSIS}
\subsection{Magnetic Field Evolution and Flare Emissions}\label{field}

The jet is located at an unipolar region near disk center with a cosine factor of 0.8 (N34$^{\circ}$W15$^{\circ}$; enclosed by the black box in Figs.~\ref{f1}$ab$), which constitutes the network on the northern boundary of a supergranule (Fig.~\ref{f1}$b$). The DAVE method yields a mean velocity of 300~m~s$^{-1}$ for the northward supergranular flows, which is of typical magnitude \citep{potts08}. The results of extrapolation models show that the jet region with positive polarity is characterized with open magnetic fields (Figs.~\ref{f1}$a$ and $c$).

\begin{figure}
\epsscale{1.17}
\plotone{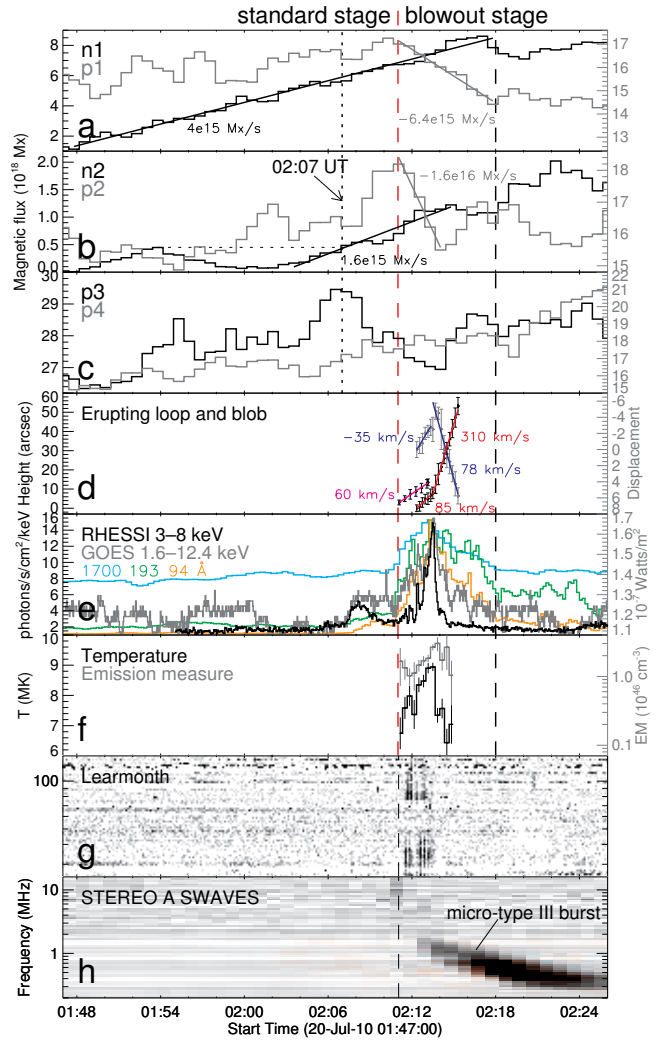}
\caption{($a$)--($c$) Temporal evolution of magnetic flux of different elements shown in Fig.~\ref{f1}. ($d$) Height-time measurement of the erupting loop and blob (see Figs.~\ref{f3} and \ref{f4}) in the projected plane. ($e$) Light curves of \hsi, \goes, and AIA (normalized). ($f$) Derived physical quantities using OSPEX. ($g$--$h$) Radio dynamic spectra from Learmonth and \stereo\ A SWAVES. \label{f2}}
\end{figure}

Magnetic flux emergence registered with the positive (p3 and p4) and negative (n1) elements (see Fig.~\ref{f1}) is seen to start from \sm01:30~UT. The evolution and flow field presented in Figures~\ref{f1}$d$--$f$ clearly show that p3 and p4 move northeast while n1 mainly moves south toward the network. Another negative element n2 emerges continuously later from 02:04~UT but does not show a significant motion. Time profiles of magnetic flux within the white boxed regions in Figure~\ref{f1}$e$ are plotted in Figures~\ref{f2}$a$--$c$. It is evident that the network elements p1 and p2 are canceled after 02:11~UT, which is caused by the collision of n1 as clearly indicated by the distance-time profile along a slit in its path (Figs.~\ref{f1}$e$ and $i$). Jet production has been known to be closely related to the cancellation of network elements with converging opposite-polarity elements \citep[e.g.,][]{wang98b,chifor08b}. Differently, however, the fluxes of n1 and n2 keep increasing during the cancellation period in this event, which can be understood if the emergence of n1 and n2 speeds up during the cancellation. Although this implies that the emerging positive flux should continuously enhance, the obviously increasing flux of p3 starts to decrease after 02:07~UT, which will be explained in \S~\ref{euv}.

Magnetic reconnection associated with the abovementioned flux cancellation is manifested as a bright, compact flare kernel over n1-p1 PIL in 1700~\AA\ images, cospatial with a \hsi\ X-ray-emitting source in 3--8~keV (Fig.~\ref{f1}$g$). This A6 microflare is initiated once n1 hits p1 at 02:11~UT (cf. Figs.~\ref{f1}$h$ and $i$) and ends at 02:18~UT when the flux cancellation ceases (Figs.~\ref{f2}$ae$). For such a small event near disk center, flare emission in the 1700~\AA\ continuum could indicate that the magnetic reconnection occurs down to heights of the temperature minimum region (TMR). Spectra fitting of the microflare using \hsi\ indicates primarily thermal X-ray bremsstrahlung emissions in the 3--8~keV energy range from a plasma of peak temperature $T=8.9\pm0.1$~MK and corresponding emission measure $EM=(2.6\pm0.3)\times10^{46}$~cm$^{-3}$ (Fig.~\ref{f2}$f$). As a comparison, microflares as small as A2 class in active regions with kG magnetic fields could bear a nonthermal component at over \sm10~keV \citep{liu04}. Nevertheless, a metric type-III storm in 25--144~MHz observed by Learmonth and a micro-type III burst in 0.2--16~MHz observed by \stereo/SWAVES are found to be co-temporal with the microflare impulsive phase (Figs.~\ref{f2}$g$$h$). These two kinds of type III emissions, indicating accelerated electrons escaping along open field lines, have been shown to result from small-scale energy release events such as jets \citep{morioka07}. We therefore speculate that this quiet-region A6 microflare did accelerate electrons to a nonthermal distribution, which was, however, overwhelmed by the instrument background of \hsi.

\subsection{Evolution of Jet} \label{euv}
\begin{figure*}
\epsscale{1.17}
\plotone{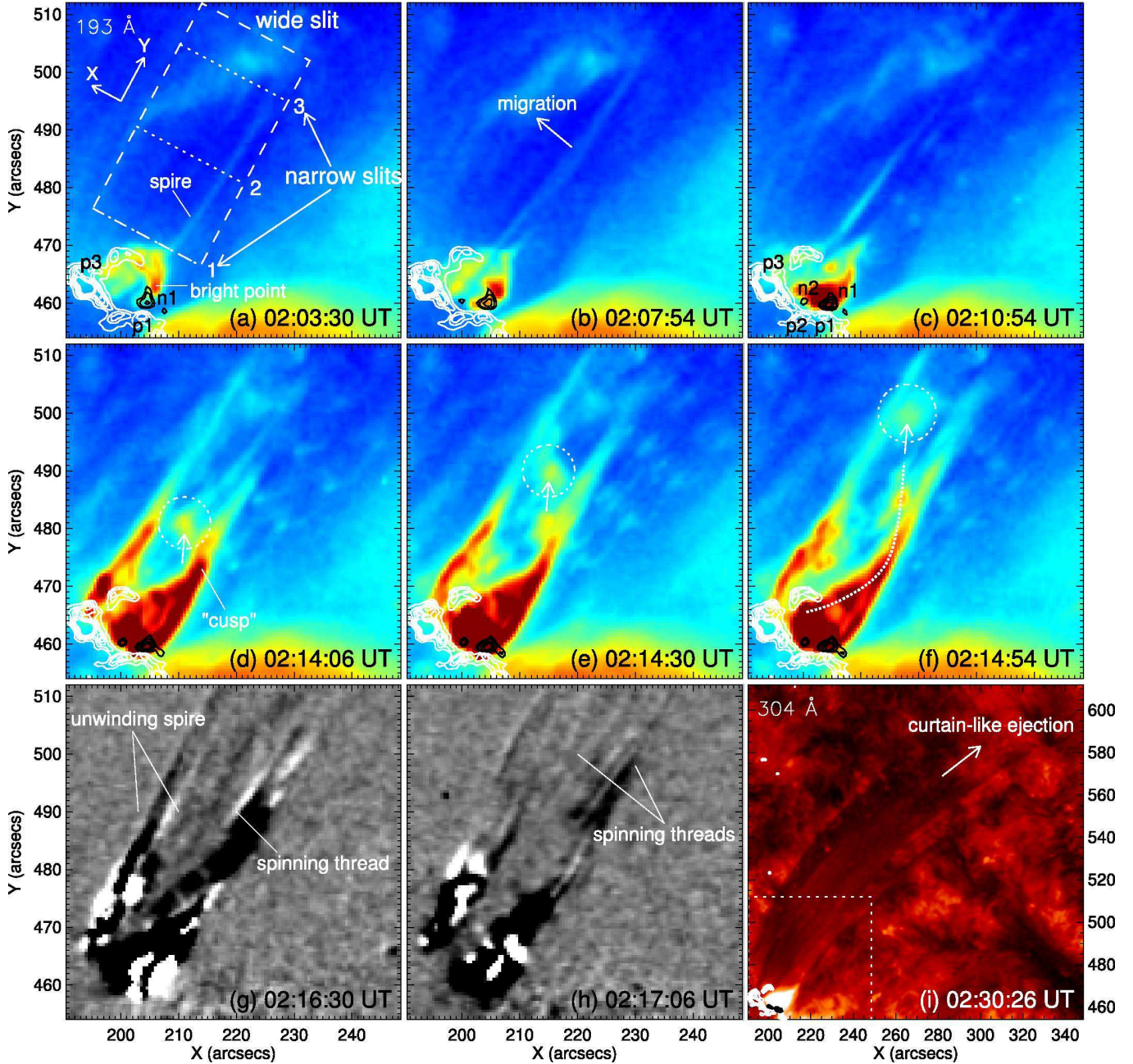}
\caption{Time sequence of EUV images. Panels $g$ and $h$ are 193~\AA\ running difference images. The dashed box in $i$ represents the FOV of other panels. The HMI contours are the same as in Fig.~\ref{f1}. The dotted circle in $d$--$f$ traces the eruption of a blob of material as in Fig.~\ref{f4}.} \label{f3}
\end{figure*}

\begin{figure}
\epsscale{1.17}
\plotone{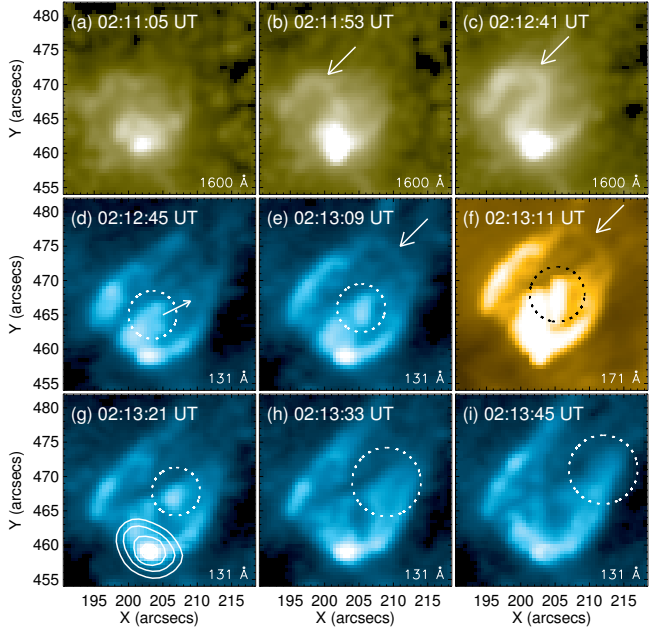}
\caption{UV/EUV images, with the arrow pointing to an erupting loop and the dotted circle tracing the eruption of a blob of material. The \hsi\ contours in $g$ are the same as in Fig.~\ref{f1}$g$.} \label{f4}
\end{figure}

Figures~\ref{f3} and \ref{f4} show representative EUV images from AIA reflecting the jet dynamics, with HMI contours superimposed on the former exhibiting the magnetic setting at the jet base. Two accompanying mpeg animations are provided to depict more dynamical detail. In Figure~\ref{f5}, distance-time profiles of a wide slit and three narrow slits (illustrated in Fig.~\ref{f3}$a$) are employed to characterize the kinematics of the ejecting material and its transverse motion in the perpendicular direction, respectively. We constructed the wide slit by orientating the long side of a 21\arcsec~$\times$~40\arcsec\ window enclosing the jet activity along the ejection direction at \sm28$^{\circ}$ clockwise from the solar north, and averaged the pixels across its short side. We then placed three narrow slits (average of a 2\arcsec\ wide, numbered 1--3) uniformly spaced by 16\arcsec\ parallel to the short side of the window, with the first one positioned at the bottom of the window. In the following, we describe the jet evolution by dividing it into two distinct stages (also denoted in Fig.~\ref{f2}).

\subsubsection{The ``Standard'' Stage}
As the emerging n1 converges toward the network from \sm01:36--02:11~UT, the jet reaches a width of \sm9~$\times 10^3$~km and a length up to \sm8~$\times 10^4$~km (in the projected plane; same as below), which are comparable to those of typical jets \citep[e.g.,][]{shimojo96}. It has a standard inverted-Y shape with a jet spire and a dome-like base enclosing a bright point above the n1-p1 PIL (Figs.~\ref{f3}$a$--$c$), which could be naturally formed following the interchange reconnection between the magnetic arch p3-n1 and the ambient open field lines rooted at p1. At least ten episodes of plasma blob stream outward along the jet spire at a speed of $\sim$300~km~s$^{-1}$ as measured by the distance-time profile of the wide slit (e.g., Fig.~\ref{f3}$c$ and Fig.~\ref{f5}$e$). As a crude estimation, if we relate the speed of n1 leading edge ($\sim$1~km~s$^{-1}$; Fig.~\ref{f1}$i$) to the reconnection inflow and the jet speed to the reconnection outflow, the reconnection rate $M$ \citep{priest00} can be estimated as $M=v_{in}/v_{out} \approx 3 \times 10^{-3}$.

A notable feature of the evolution is that from 02:07~UT, the jet spire migrates northeastward away from the n1-p1 PIL (cf. Figs.~\ref{f3}$a$--$c$) at a speed of 8~km~s$^{-1}$ (projected speed; same as below) measured by the slit 1 at the top of the dome (see Figs.~\ref{f3}$a$ and \ref{f5}$a$). In the meantime, the magnetic element n2 continuously emerges beyond its previous state (Fig.~\ref{f2}$b$) and by \sm02:11~UT, the bright point extends to encompass n2 forming a more complete circular dome (Fig.~\ref{f3}$c$). These evince that the newly emerged field p3-n2 speeds up the interchange reconnection and the strengthened, newly formed open field lines rooted at p3 are released upward by the magnetic tension force. We note that although the potential field extrapolation does not demonstrate well the connectivity between p3 and n1/n2, the model result largely matches the jet features of topological interest (cf. Fig.~\ref{f1}$c$ and Fig.~\ref{f3}$c$). Spires of polar coronal jets are generally observed to migrate laterally at a comparable speed during the maximum phase \citep{savcheva07,savcheva09} and is featured in the standard jet model \citep{moore10}. It is, however, not found before 02:07~UT when there seems to have weaker reconnection. In this regard, our result of weak to strong reconnection corresponding with fixed to migrating jet spire is in line with that of the quasi-static to eruptive evolution of the jet material bundle \citep{liu+wei+jet09} and that of the slow to fast reconnection in jet simulations \citep{pariat10}.

As the jet spire rooted at p3 swings toward the pole away from the observer, the measured surface field of the element p3 should decrease due to change of the field line orientation \citep[e.g.,][]{liu05,wang10}. This explains the decrease of the p3 flux after 02:07~UT despite of the flux emergence as discussed in \S~\ref{field}. Conversely, we may reason that p4 is not involved in the jet reconnection as discernible in EUV images (Figs.~\ref{f3}$c$--$f$), since its flux does not diminish after 02:07~UT (Fig.~\ref{f2}$c$).

\subsubsection{The ``Blowout'' Stage}
The succeeding collision of n1 and p1 producing the microflare from 02:11~UT is apparently followed by a decrease of the converging speed (Fig.~\ref{f1}$i$), which is most probably due to the pileup of magnetic flux \citep{chae02}. As the cancellation occurs at a rate of $R \approx 6.4 \times 10^{15}$~Mx~s$^{-1}$ (Fig.~\ref{f2}$a$) along the contact length $l \approx 2.6 \times 10^{8}$~cm (Fig.~\ref{f1}$e$), the cancellation rate per unit length is $r=R/l=2.5 \times 10^7$~G~cm~s$^{-1}$. The flux pileup reconnection model of \citet{litvinenko07} has established a linkage between $r$ and the physical properties of the Sweet-Parker current sheet, which yields $v_{in} \approx 330$~m~s$^{-1}$ and $v_{out} \approx 8$~km~s$^{-1}$ for this event in the TMR. Therefore, the reconnection rate $M \approx 0.04$ is significantly larger than that in the ``standard'' stage, which could point to a more effective removal of the outer fields of the emerging flux to facilitate a subsequent, faster emergence as inferred in \S~\ref{field}. We also conjecture that the reconnection involving p2 that does not directly collide with n1 could occur higher, which might explain the lack of emission in 1700~\AA\ but brighter EUV emission above p2 (Figs.~\ref{f3}$d$--$f$). We further note the following. 

\begin{figure}
\epsscale{1.17}
\plotone{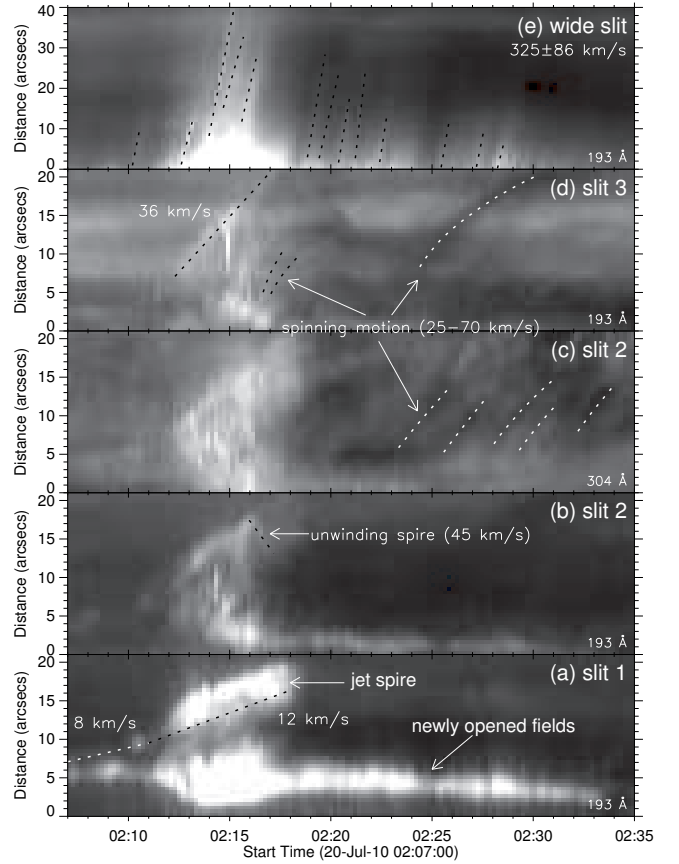}
\caption{Time slices for the slits 1--3 ($a$--$d$) and the wide slit ($e$) depicted in Fig.~\ref{f3}$a$. The distance is measured following the $+X$ and $+Y$ direction in Fig.~\ref{f3}$a$ for the narrow and wide slits, respectively. \label{f5}}
\end{figure}

First, the jet spire becomes three times wider and continuously brightened, and comparing with the stage 1, it starts to swing at a faster speed of 12--36~km~s$^{-1}$ (see Figs.~\ref{f3}$d$--$h$ and \ref{f5}$ad$), which manifests the enhanced flux emergence and reconnection. Interestingly, when looking from above toward the jet base, the spire conspicuously unwinds itself counterclockwise for at least half a turn during 02:16--02:18~UT before it vanishes with the ending of the microflare. This is clearly exhibited by its helical morphology (the alternating white and black features in Figs.~\ref{f3}$gh$) and an inward drifting structure in the slit image yielding an apparent transverse velocity of 45~km~s$^{-1}$ (Fig.~\ref{f5}$b$). The unwinding jet spire strongly points to the presence of left-handed magnetic twist in the emerging magnetic fluxes, which can be transferred to the newly formed open fields via the interchange reconnection and then propagates upward like a torsional Alfv\'{e}n wave \citep{shibata85,shibata86}. The unwinding motion persists for such a short duration that we are unable to derive its propagating velocity (however, see \citealt{liu+wei+jet09}).

Second, a loop and a blob are seen to rise and erupt during the flare impulsive phase. The loop appears to connect p3 and n1/n2 and is most discernible in cooler EUV channels (Figs.~\ref{f4}$a$--$c$ and $e$--$f$). In a rotated frame with the direction along the long side of the wide slit as $y$-axis (see Fig.~\ref{f3}$a$), the loop front travels outward at $v_y=60$~km~s$^{-1}$ (Fig.~\ref{f2}$d$), under which a bright blob, discernible in all the AIA EUV bands, emerges and also erupts outward. The blob feature (the dashed circle in Fig.~\ref{f4}$d$--$i$ and Fig.~\ref{f3}$d$--$f$) first moves at $(v_x, v_y)=(-35,85)$~km~s$^{-1}$ then accelerates to $(v_x, v_y)=(78,310)$~km~s$^{-1}$ after the flare peaking at 02:13:30~UT (Figs.~\ref{f2}$de$).
Its path (the dotted line in Fig.~\ref{f3}$f$) seems to correspond to a two-dimensional projection of a three-dimensional spiral motion with a counterclockwise rotation, considering that the nearby jet spire also shows a coherent spinning motion subsequently. Remarkably, the loop-blob structure at \sm02:13:10~UT (Figs.~\ref{f4}$ef$) is reminiscent of the front-core structure of typical CMEs \citep[e.g.,][]{illing86}. We surmise that our observed erupting loop and blob might be a miniature version of CMEs as speculated by \citet{moore10}, in which the blob could be the embedded dense material representing the twisted core of the escaping magnetic fluxes. Additional support to this inference comes from the cusp-shaped bright loop above the extended bright points n1/p1 and n2/p2 (Figs.~\ref{f4}$i$ and \ref{f3}$d$--$h$). The cusp is formed right after the ejection of the loop/blob, which most probably stretch open the closed fields of the bright points to allow the hot plasma to stream outward (see Figs.~\ref{f5}$ae$). By this stage, it is obvious that the dome structure is completely blown out.

Third, all the jetting activities exhibit a velocity of $325\pm86$~km~s$^{-1}$ (Fig.~\ref{f5}$e$), comparable to the jet speed in other events obtained using spectroscopy \citep[e.g.,][]{moreno08}. The ejecting material observed in chromospheric and transition-region lines forms a multi-stranded curtain-like structure (Figs.~\ref{f3}$g$--$i$), which is similar to those found in the polar blowout-type jets \citep{moore10} and bolsters the view of the blowout dome. Moreover, those jets threads also rotate counterclockwise bearing a transverse velocity of 25--70~km~s$^{-1}$ (Figs.~\ref{f5}$cd$), suggesting the unwinding of left-handed twisted magnetic fluxes as they move into the upper atmosphere.

\section{CONCLUSION}
Combining the analysis of photospheric magnetic field evolution and coronal dynamics, we have presented a disk jet with the occurrence of a microflare marking the transition from a standard to a blowout jet activity. Such a study of small-scale energy release event is made possible thanks to the unprecedented data from \sdo. To our best knowledge, this is the first detailed observation of the recently proposed blowout-type jets \citep{moore10}. To link up the observational results, we suggest that the jet first proceeds in a standard fashion. Then with the sudden occurrence of flux cancellation, a microflare in the TMR sets in, which rapidly and sufficiently weakens the tension of the outer envelope field of the emerging flux, and as a result the twisted core field is unleashed, transferring the twist to the ambient, opening field lines via interchange reconnection, and blowing open the dome of the jet.

The above scenario is apparently analogous to the breakout process involved in the production of major CMEs \citep{antiochos99}. Our observations of the erupting loop and blob, which were closely associated with the jet evolution, provide direct support for the model of \citet{moore10}, which depicts a jet-scale magnetic breakout. More high-resolution observations of disk jets with both coronal and magnetic measurements available are desired in order to further understand the physics of the jet phenomenon.

\acknowledgments
\sdo\ is a mission for NASA's Living With a Star program. \hsi\ is a NASA Small Explorer. We thank the referee for valuable comments. C.L., R.L., S.W., and H.W. were supported by NSF grants AGS 08-19662 and AGS 08-49453, and NASA grants NNX 08AQ90G and NNX 08AJ23G. N.D. was supported by NASA grant NNX 08AQ32G.

\end{document}